\documentclass[showpacs,prl,preprint,superscriptaddress]{revtex4}

\usepackage{amsmath}
\usepackage{graphicx}
\usepackage{dcolumn}
\usepackage{bm}

\begin{document}
\title{Entanglement, BEC, and superfluid-like behavior of two-mode photon systems}
\author{Ferdinando  de Pasquale}
\affiliation{Dipartimento di Fisica, Universit\`{a} di Roma La
Sapienza, Piazzale A. Moro 2, 00185 Roma, Italy} \affiliation{INFM Center for Statistical Mechanics and Complexity, Piazzale A. Moro 2, 00185 Roma, Italy}
\author{Gian Luca Giorgi}
\email{gianluca.giorgi@roma1.infn.it}\affiliation{INFM Center for Statistical Mechanics and Complexity, Piazzale A. Moro 2, 00185 Roma, Italy}
\affiliation{Dipartimento di Fisica, Universit\`{a} di Roma La
Sapienza, Piazzale A. Moro 2, 00185 Roma, Italy}

\begin{abstract}
A system of two interacting photon modes, without constraints on the photon number, in the presence of a Kerr nonlinearity, exhibits BEC if the transfer amplitude is greater than the mode frequency.
A symmetry-breaking field (SBF) can be introduced by taking into account a classical electron current. The ground state, in the limit of small nonlinearity, becomes a squeezed state, and thus the modes become entangled. The smaller is the SBF, the greater is entanglement. Superfluid-like behavior is observed in the study of entanglement growth from an initial coherent state, since in the short-time range the growth does not depend on the SBF amplitude, and on the initial state amplitude. On the other hand, the latter is the only parameter which determines entanglement in the absence of the SBF.
%The occupation number correlations are studied for an initial coherent state both in the BEC and in the superfluid case. Results are compared with the macroscopic approximation, where fluctuations of the field operators are neglected.

\end{abstract}
\maketitle

Quantum entanglement is considered as a fundamental resource in quantum
information science \cite{nielsen}. In the last years, quantum information
with continuous variables (CVs) \cite{braunstein,illuminati} has receveid a
lot of attention because of the simplicity of preparing, unitarily
manipulating, and measuring quantum states. Various CV applications have
been considered such as teleportation \cite{kimble} and cryptography \cite
{critt}.

At the same time, quantum phase transitions (QPTs) are recognized as a
fundamental concept in quantum many-body systems \cite{sachdev}. In
particular, the transition to a ground state with a macroscopic occupation,
the Bose-Einstein condensation (BEC) regime, is widely studied both
theoretically and experimentally \cite{davis,bradley,anderson}. Recently, a
great effort has been devoted in understanding how quantum correlations
(such as entanglement) are related to the critical behavior of systems
exhibiting QPT effects \cite{amico,osborne}.

Remarkable examples of QPTs in CV systems are superradiance \cite
{dicke,lambert} and Bose-Einstein condensation (BEC) in microcavity
polaritons \cite{weisbuch}. Some years ago, Chiao investigated the
possibility of observing BEC and superfluidity in photon gases \cite{chiao}.
In fact, since photon quantized in a three-dimension cavity are massless and
their chemical potential is zero, BEC transition would appear forbidden. To
overcome this difficulty, Chiao proposed a two-dimensional array of
Fabry-Perot cavities where an effective chemical potential can be defined
due to quantization conditions. A similar approach has been also proposed by Navez \cite{navez}.

In this paper we show that a ground state with a finite occupation (BEC) for
pure photonic systems is indeed possible without the need of special
architectures. We will consider two photon modes, coupled via an exchange
interaction, in the presence of a weak nonlinearity. The emergence of an
effective ``chemical potential'', manifests itself as a consequence of the
unitary transformation which diagonalizes the quadratic part of the
Hamiltonian. The normal phase corresponds to the situation where the photon
frequency exceeds the interaction coupling, while in the opposite case a BEC
phase comes out. In the anomalous phase, a finite amount of entanglement in
the ground state is predicted between the two modes which grows with the
occupation number. In Ref. \cite{lambert} a two-mode boson entanglement
across the transition from a normal to a superradiant phase is discussed.
Squeezing and entanglement in the ground state are present in both phases.
The main difference with our model is that squeezing and entanglement appear
only \ in the anomalous phase. Other several model, discussing the
interaction of two e.m. modes in nonlinear media have been also considered
\cite{imoto,gerry}.
While ground-state properties allow the investigation of the quantum nature of the system, quantum optics experiments are usually set to observe the transient regime determined by cavity losses \cite{giacobino}. Then, it seems important the discussion of how an initially factorized state can develop entanglement in its time evolution. The evolution of entanglement
from a given initial state has been studied in Ref. \cite{sanz} for Fock and
coherent states. In the cases considered there, the evolution does not
depend on the presence of BEC. On the other hand, if a symmetry-breaking field (SBF), physically achievable through the interaction of the e.m. radiation
with a classical electron current is present, a ``superfluid-like'' phase appears. % In this case the dynamical generation of entanglement from a
%product of coherent states is quite different in the normal and in the anomalous phase.
The main difference with the work of Ref. \cite{sanz} is the existence of an initial time range where the entanglement growth does not depend on the SBF amplitude.

Let us start with the description of the model by considering two coupled photon systems $a_{i}$ ($i=1,2$) with the same
energy frequency $\omega $, a transfer interaction of strength $w$, and a
nonlinear Kerr interaction of strength $g$
\begin{equation}
H=\omega \left( a_{1}^{\dagger }a_{1}+a_{2}^{\dagger }a_{2}\right) -w\left(
a_{1}^{\dagger }a_{2}+a_{2}^{\dagger }a_{1}\right) +g\left(
n_{1}+n_{2}\right) ^{2}
\end{equation}
The two systems could be associated to the polarization components of the
same e.m. field. This is the case where the assumption about the interaction
is sensible.

To diagonalize $H$\ we perform a rotation on the original degrees of freedom
$a_{i}$ which defines new variables $\alpha =\left( a_{1}+a_{2}\right) /%
\sqrt{2}$ and $\beta =\left( a_{1}-a_{2}\right) /\sqrt{2}$.

This unitary transformation cancels the transfer term\qquad\
\begin{equation}
H=\left( \omega -w\right) \alpha ^{\dagger }\alpha +\left( \omega +w\right)
\beta ^{\dagger }\beta +g\left( \alpha ^{\dagger }\alpha +\beta ^{\dagger
}\beta \right) ^{2}.  \label{hamil}
\end{equation}
We observe that the last term, even if small, becomes important for $%
w>\omega $, when the vacuum photon state is no more the ground state. In
fact, in the absence of interaction, the Hamiltonian is no more bounded from
below.

\begin{figure}[htbp]
\begin{center}
\includegraphics{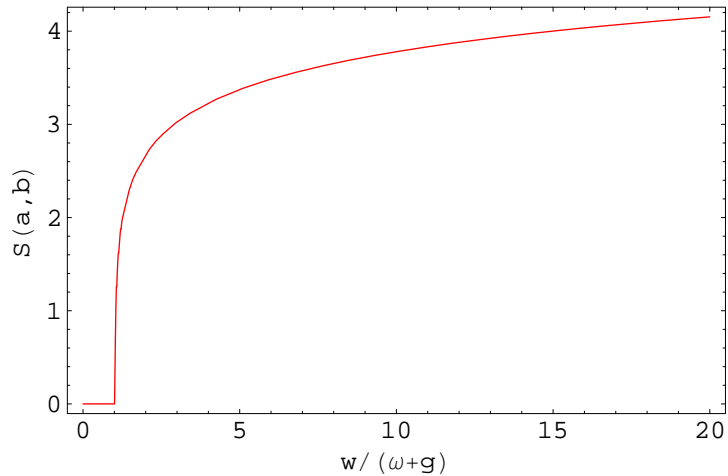}
\end{center}
\caption{Von Neumann entropy of the subsystems $a$ and $b$ versus the critical parameter $w/(\omega+g)$. In the normal phase no entanglement is detected, while beyond the critical point the ground state is no longer factorizable. The system's parameter are $\omega=1$ and $g=0.01$. } \label{fig:fig1}
\end{figure}

The Hamiltonian (\ref{hamil}) is diagonal in the basis spanned by the Fock
states in the $(\alpha,\beta) $ representation \cite{sanz}. The ground state
is the vacuum when $\left( \omega -w\right) >0$, whereas it becomes\ $\left|
n_{\alpha }\right\rangle \left| 0_{\beta }\right\rangle $ when $\left(
\omega -w\right) <0$ , where $n_{\alpha }$ is the integer closest to $\left(
w-\omega \right) /\left( 2g\right) $. In the latter case the corresponding
eigenvalue is $\left( \omega -w\right) n_{\alpha }+gn_{\alpha }^{2}$. Ground state properties are remarkably different in the two cases.
Studying the entanglement of formation between the modes $a$ and $b$ by
means of the Von Neumann entropy $S$ of either of the reduced states, two
regimes appear. In the normal phase, when the ground state is the vacuum,
there is no entanglement, while when the ground state is $\left| n_{\alpha
}\right\rangle \left| 0_{\beta }\right\rangle $ the following expression
for $S$ holds
\begin{equation}
S=-\frac{1}{2^{n_{\alpha }}}\sum_{k=0}^{n_{\alpha }}
{n_{\alpha } \choose k}
\log \left[ \frac{1}{2^{n_{\alpha }}}
{n_{\alpha } \choose k}\right] ,
\end{equation}
which shows how the Von Neumann entropy grows with $n_{\alpha }$. In Fig. \ref{fig:fig1}
the behavior of $S(a,b)$ is plotted as a function of the critical parameter $w/(\omega+g)$.

As said, ground state properties are significant, but more important can be the
achievement of an experimental method to create entanglement and to exploit
it in quantum information protocols. The authors of Ref. \cite{sanz} have
studied the entanglement generation induced in the time evolution of Fock
states and coherent state. Let us focalize our attention on these latter
states. We note that the dynamical entanglement does not depend on the
presence or the absence of the BEC phase. This result can be understood by
observing that $\omega$ can be dropped through the introduction of a ``local'' unitary
transformation which cannot determine any change in the entanglement value.
Furthermore, the amplitude of the incoming coherent state influences deeply
the degree of entanglement. Then, low intensity two-mode laser fields are
expected to develop a little amount of entanglement. The introduction of a finite SBF modifies these
properties in a significant way. Let consider a field of amplitude $\lambda
/\sqrt{2}$ acting on both the modes $a$ and $b$\ . The new Hamiltonian is $%
H_{\lambda }=H-\lambda \left( \alpha ^{\dagger }+\alpha \right) $. This
field does not represent a purely mathematical tool introduced to describe the
emergence of a superfluid phase, as usual in boson particles systems, but it can be
physically realized through the interaction of the e.m. modes with a
``classical'' non-fluctuating electron current \cite{glauber}, and, then, it
can be kept finite. An exact solution for this model does not exist anymore,
and a perturbation theory is worth to be developed. First, in order to study
the properties of the ground state of $H$, it is easy to recognize that the
vacuum state for the mode $\beta $ minimizes the energy. Then, we write the
ground state $\left| G\right\rangle $ as the product $\left| G_{\alpha
}\right\rangle \left| 0_{\beta }\right\rangle $. Let us try to do a
perturbation theory starting from the solution known for $\lambda =0$. In
the normal phase the correction to the ground state energy is $\Delta
E=-\lambda ^{2}/\left( \omega -w+g\right) $. The result is substantially
different in the anomalous phase. When $\left( w-\omega \right) /2g$ is an
integer, the first excited states are degenerate ($\left| n_{\alpha
}+1\right\rangle $ and $\left| n_{\alpha }-1\right\rangle $\ have the same
energy) and the energy gap is $g$: $\Delta E=-\lambda ^{2}\left( 2n_{\alpha
}+1\right) /g$. If $\left( w-\omega \right) /2g$ is not integer, the solution is just slightly different. A qualitatively different solution arise when $\left( w-\omega \right) /2g$ falls exactly between two integer numbers. Now the ground state is degenerate, and energy corrections are of order $ \lambda$. Apart from these special values, in the limit of small nonlinearity, which corresponds to $g<\lambda $, the energy spreading
is greater than the gap. Then, the system becomes unstable with respect to
this perturbation, and a different approach is in order.

Next, we show that the effect of nonlinearity can be taken into
account introducing a ``superfluid-like'' phase in the limit of small $g$.
We mean that in the new ground state the annihilation operator $\alpha $
has a nonvanishing mean value $\nu $. It is then natural to introduce a
translation of amplitude $\nu $ ($\alpha \rightarrow \alpha +\nu $). The
Hamiltonian becomes $H_{\lambda }=E_{0}+H_{1}+H_{2}+H_{3}+H_{4}$, where
\begin{eqnarray}
E_{0} &=&\left( \omega -w+g\right) \nu ^{2}+g\nu ^{4}-2\lambda \nu , \\
H_{1} &=&\left( \alpha ^{\dagger }+\alpha \right) \left[ \left( \omega
-w+g\right) \nu +2g\nu ^{3}-\lambda \right] , \\
H_{2} &=&\left( \omega -w+g\right) n_{\alpha }+\left( \omega +w+g\right)
n_{\beta }+g\nu ^{2}\left[ 4n_{\alpha }+ \alpha ^{\dagger 2}+\alpha
 ^{2}\right] , \\
H_{3} &=&2g\nu \left[ \left( \alpha ^{\dagger }+\alpha \right) n_{\beta
}+\alpha ^{\dagger }n_{\alpha }+n_{\alpha }\alpha \right] , \\
H_{4} &=&g\left[ \alpha ^{\dagger 2}\alpha ^{2}+\beta ^{\dagger 2}\beta
^{2}+2n_{\beta }n_{\alpha }\right] .
\end{eqnarray}

The weak-coupling approximation amounts to take into account terms of order $%
g\nu ^{2}$, and to neglect terms of order $\sqrt{g}$ and $g$. In this limit
we can disregard \ $H_{3}$ and $H_{4}$. The condensate amplitude $\nu $ can
be fixed by minimizing $E_{0}$:
\begin{equation}
\left( \omega -w+g\right) \nu +2g\nu ^{3}-\lambda =0.  \label{nnumin}
\end{equation}
We note that this conditions makes $H_{1}$ vanishing.

%Let consider $\lambda $ finite but small. If we observe $E_{0}$ and compare
%it with the energy of the ground state for $\lambda =0$, that is with $%
%\left( \omega -w\right) n_{\alpha }+gn_{\alpha }^{2}$, two different
%situation when

Two different behaviors appear depending on the sign of the coefficient of $%
\nu $. If $\left( \omega -w+g\right) >0$ we have $\nu \simeq \lambda /\left(
\omega -w+g\right) $. In this case we have a ``normal'' behavior, in the
sense that condensate amplitude vanishes with $\lambda $. On the other hand,
for $\left( \omega -w+g\right) <0$, a solution for the minimum of $%
E_{0}\left( \nu \right) $ is given by $\nu \simeq \nu ^{\ast }+\lambda
/\left( 4g\nu ^{\ast 2}\right) $, with $\nu ^{\ast }=\sqrt{\frac{w-\omega -g%
}{2g}}$. This solution corresponds to a superfluid state. %It could be
%achieved by switching on an intense field $\lambda $, and then by
%adiabatically driving it to a value close to zero. According to the
%weak-coupling approximation, this result holds when $\lambda $ is assumed to
%be greater than $g$ but also $\lambda /\left( 4g\nu ^{\ast 2}\right) \ll \nu
%^{\ast }$, that is $\lambda \ll \sqrt{\frac{\left( w-\omega -g\right) ^{3}}{g%
%}}$.

Taking into account Eq.(\ref{nnumin}), we obtain for the Hamiltonian $ H_2=H_{\alpha }+H_{\beta }$, where
\begin{eqnarray}
%H &=&H_{\alpha }+H_{\beta } \\
H_{\alpha } &=&\frac{\lambda }{\nu }n_{\alpha }+g\nu ^{2}\left(2n_{\alpha }+ \alpha
^{\dagger 2}+\alpha ^{2}\right)   \label{accaalfa} \\
H_{\beta } &=&\left( \omega +w+g\right) n_{\beta }
\end{eqnarray}
Qualitatively, we observe that in the normal phase $H_{\alpha }$ is
dominated by the term $\left( \lambda /\nu \right) n_{\alpha }$\ and the
ground state is expected to be very close to the vacuum state. Conversely, in the
anomalous phase a squeezed state on the mode $\alpha $ will appear. Thus,
entanglement is expected to come out between the modes $a$ and $b$, in
analogy with the well known effect concerning the passage of squeezed light
through a beam-splitter.

It is worth noting that for $\lambda =0$ a quadratic dependence on the
operator $x_{\alpha }=\left( \alpha ^{\dagger }+\alpha \right) /\sqrt{2}$
comes out. Thus $x_{\alpha }$ is a constant of motion, and the canonical
conjugate variable $p_{\alpha }=i\left( \alpha ^{\dagger }-\alpha \right) /%
\sqrt{2}$ freely diffuses. This feature has been recognized in Bose
particles systems by Lewenstein \cite{lewenstein}. It is commonly assumed
(see for a review \cite{walls}) that corrections of order $g$ are able to
remove this singular behavior. %As we will discuss in a forthcoming paper, a self-consistent treatment of correction of order $\sqrt{g}$ does not affect the presence of the singularity. The robustness of this result seems to be related to a generalization of the Goldstone theorem.
This result is known in the Bose particles system as the Hartree-Fock-Bogoliubov approximation \cite{griffin}, and rejected on the basis of general arguments (Goldstone theorem).

For finite, even if small $\lambda $, it exists a canonical transformation
(the Bogoliubov transformation) which diagonalizes $H_{\alpha }$. If we
define $\gamma =\cosh \theta \alpha -\sinh \theta \alpha ^{\dagger }$, we
get $H_{\alpha }=\epsilon \gamma ^{\dagger }\gamma +\epsilon _{0},$ where $%
\epsilon =\sqrt{\frac{\lambda }{\nu }\left( \frac{\lambda }{\nu }+4g\nu
^{2}\right) }$ and $\epsilon _{0}$ is the zero-point energy. The parameter $%
\theta $ is defined by $\tanh 2\theta =-2g\nu ^{2}/\left( \frac{\lambda }{%
\nu }+2g\nu ^{2}\right) $. The ground state $\left| G_{\alpha }\right\rangle
$ of $H_{\alpha }$ is the squeezed state $\left| G_{\alpha }\right\rangle
=\exp \left[ \frac{\theta }{2}\left( \alpha ^{\dagger }\alpha ^{\dagger
}-\alpha \alpha \right) \right] \left| 0\right\rangle $. In the original
space $a,b$, taking into account of the translation performed on $\alpha $,
\begin{equation}
\left| G_{a,b}\right\rangle =\exp \left[ \frac{\theta }{4}\left( a^{\dagger
}a^{\dagger }+b^{\dagger }b^{\dagger }+2a^{\dagger }b^{\dagger }-\sqrt{2}\nu
\left( a^{\dagger }+b^{\dagger }\right) \right) -h.c.\right] \left|
0,0\right\rangle .
\end{equation}
In the Schr\"{o}dinger representation, where the coordinates are $%
x_{a}=\left( a^{\dagger }+a\right) /\sqrt{2}$ and $x_{b}=\left( b^{\dagger
}+b\right) /\sqrt{2}$, the wave function reads as
\begin{equation}
\Psi _{G}\left( x_{a},x_{b}\right) =\exp \left\{ -\frac{e^{-2\theta }}{4}%
\left[ \left( x_{a}+x_{b}\right) +2\nu \right] ^{2}-\frac{1}{4}\left(
x_{a}-x_{b}\right) ^{2}\right\} .
\end{equation}
Given the wave function, and then a complete description of the state, we
characterize the phase transition through the entanglement of formation
between the subsystems $a,b$. Since we have a pure two-mode state, the Von
Neumann entropy of either of the reduced states is a unique entanglement
measure. From our wave function, the Von Neumann entropy is \cite{rendell}
\begin{equation}
S\left( a,b\right) =\cosh ^{2}\frac{\theta }{2}\log \left( \cosh ^{2}\frac{%
\theta }{2}\right) -\sinh ^{2}\frac{\theta }{2}\log \left( \sinh ^{2}\frac{%
\theta }{2}\right) .
\end{equation}

\begin{figure}[htbp]
\begin{center}
\includegraphics{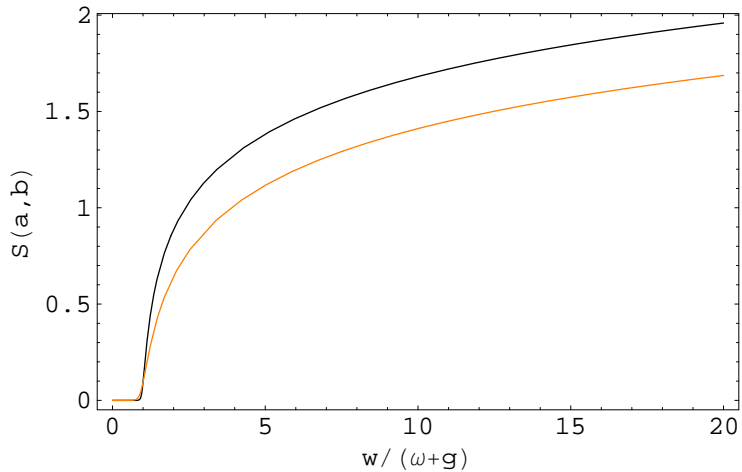}
\end{center}
\caption{Von Neumann entropy of the subsystems $a$ and $b$ versus the critical parameter $w/(\omega+g)$ in the presence of a SBF $\lambda$. The system's parameter are $\omega=1$ and $g=0.01$. The orange (light gray) line refers to $\lambda=0.3$,
   while the black (dark gray) line corresponds to $\lambda=0.1$ } \label{fig:fig2}
\end{figure}

In Fig. \ref{fig:fig2} $S\left( a,b\right) $ is plotted as a function of the critical
parameter $w/\left( \omega +g\right) $ for different values of $\lambda $.
In the normal phase there is not entanglement, while in the condensate phase
$S\left( a,b\right) $ becomes drastically different from zero. The plot
shows that the smaller is $\lambda $ the bigger is the amount of
entanglement in the ground state.

Coming back to the problem represented by a dynamical generation of
entanglement starting by a factorized state, we next consider the dynamics
of a coherent state $\left| \Phi \right\rangle =\left| \nu ^{\prime },\nu
^{\prime }\right\rangle $ with the same amplitude $\nu ^{\prime }$ on both
the modes $a$ and $b$. Furthermore, we assume that the system is in the
condensate phase. In the space of $\alpha $ and $\beta $ this state
corresponds to the coherent state of amplitude $\nu ^{\prime }\sqrt{2}$ for
the mode $\alpha $ times the vacuum for $\beta $. To solve the problem of
calculating $e^{-iHt}\left| \Phi \right\rangle $ we use the trivial
identities $\alpha \left( -t\right) \left| \Phi \left( t\right) \right\rangle  =\nu
^{\prime }\sqrt{2}\left| \Phi \left( t\right) \right\rangle$ and $\beta \left( -t\right) \left| \Phi \left( t\right) \right\rangle  =0$.
%\begin{eqnarray*}
%\alpha \left( -t\right) \left| \Phi \left( t\right) \right\rangle  &=&\nu
%^{\prime }\sqrt{2}\left| \Phi \left( t\right) \right\rangle  \\
%\beta \left( -t\right) \left| \Phi \left( t\right) \right\rangle  &=&0
%\end{eqnarray*}
The evolution of the annihilation operators in the Heisenberg picture can
be obtained noting that $\gamma \left( t\right) =\exp \left( -i\epsilon
t\right) \gamma $ and $\beta \left( t\right) =\exp \left[ -i\left( \omega
+w+g\right) t\right] \beta $. After the substitution of these expressions
and after some algebraic manipulations, the following result is established:
\begin{equation}
\alpha \left( t\right) =f\left( t\right) \alpha +f^{\prime }\left( t\right)
\alpha ^{\dagger }+h\left( t\right) ,
\end{equation}
%and, analogously,
%\begin{equation}
%b\left( t\right) =f\left( t\right) b+f^{\prime }\left( t\right) a+\left(
%a^{\dagger }+b^{\dagger }\right) f^{\prime \prime }\left( t\right) +h\left(
%t\right) ,
%\end{equation}
where
\begin{eqnarray}
f\left( t\right)  &=&\cos \epsilon t-i\cosh 2\theta \sin \epsilon t,
\label{ae1} \\
f^{\prime }\left( t\right)  &=&i\sinh 2\theta \sin \epsilon t,  \label{ae2}
\\
h\left( t\right)  &=&\nu -\nu \left( \cos \epsilon t-ie^{-2\theta }\sin
\epsilon t\right) .  \label{ae3}
\end{eqnarray}
%In the limit of $\lambda \rightarrow 0$, since $e^{-2\theta }\sin \epsilon
%t\rightarrow 4g\nu ^{2}t$, these solutions give rise to the phase diffusion
%effect \cite{lewenstein}.

\begin{figure}[htbp]
\begin{center}
\includegraphics{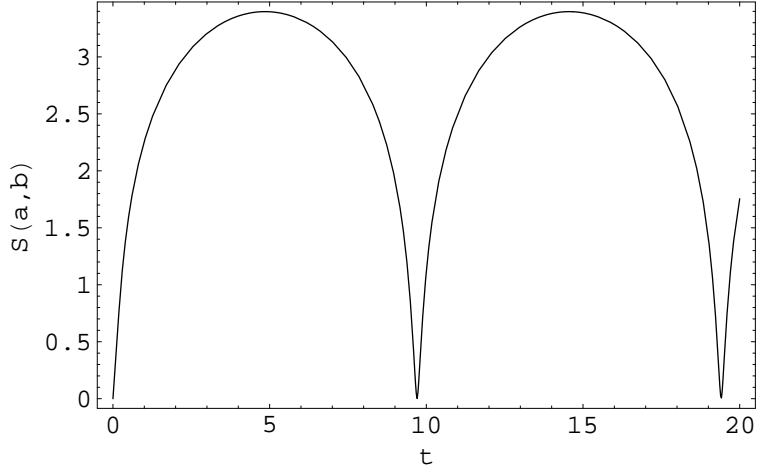}
\end{center}
\caption{Dynamical evolution of entanglement as a function of time. The initial state is the product of two coherent states on the modes $a$ and $b$. The system's parameter are $\omega=1$, $w=2$, $g=0.1$, and $\lambda=0.11$. } \label{fig:fig3}
\end{figure}

%From Eqs.(\ref{ae1},\ref{ae2},\ref{ae3})
Through these results, two coupled partial differential
equations can be written for the wave function $\Psi \left(
x_{a},x_{b}\right) =\left\langle x_{a},x_{b}|\Phi \left( t\right)
\right\rangle $ which allow to solve the problem. The equations are
\begin{eqnarray}
\left[ \left( f+f^{\prime }\right) \left( x_{a}+x_{b}\right) +\left(
f-f^{\prime }\right) \left( \frac{\partial }{\partial x_{a}}+\frac{\partial
}{\partial x_{b}}\right) +\sqrt{2}\left( h-\nu ^{\prime }\right) \right]
\Psi \left( x_{a},x_{b},-t\right)  &=&0 \\
\left( x_{a}-x_{b}+\frac{\partial }{\partial x_{a}}-\frac{\partial }{%
\partial x_{b}}\right) \Psi \left( x_{a},x_{b},t\right)  &=&0\qquad
\end{eqnarray}
Apart from a normalization constant, the solution is
\begin{eqnarray}
\Psi \left( x_{a},x_{b},-t\right)  &=&\exp \left[ -\frac{h-\nu ^{\prime }}{%
f-f^{\prime }}\left( x_{a}+x_{b}\right) \right]   \nonumber \\
&&\times \exp \left[ -\frac{1}{2}\left( \frac{f}{f-f^{\prime }}\right)
\left( x_{a}^{2}+x_{b}^{2}\right) +\left( \frac{f^{\prime }}{f-f^{\prime }}%
\right) x_{a}x_{b}\right]
\end{eqnarray}
The last term is responsible for entanglement, and is intrinsically related
to the squeezing effect. In fact, it would be zero for $\theta =0$. The
initial amplitude $\nu ^{\prime }$ does not play any role in the
entanglement evolution. The Von Neumann entropy (whose analytical expression
is rather intricate) is plotted in Fig. \ref{fig:fig3}  as a function of time.

An oscillation with frequency $2\epsilon $ is observed. As a crucial point of
this derivation, we remark that the entanglement is completely independent
from the amplitude of the incoming coherent state. This result is a specific
signature of the superfluid phase. In fact, in the absence of the
SBF, the evolution of a coherent state involves an
amount of entanglement which is strongly related to its amplitude and goes
to zero as $\nu ^{\prime }$ does. Further, the short-time evolution shows a weak dependence on the SBF amplitude. The limit of validity of the free-diffusion regime is given by $\lambda\ll2g\nu^3$ together with $\epsilon t\ll1$. %Then, we propose to observe experimentally
%the emergence of the superfluid phase by pumping the nonlinear cavity with a
%low-intensity laser field. In this regime we expect a finite amount of
%entanglement, in contrast with what would happen in the normal phase, where
%the results should be similar to those drawn in Ref. \cite{sanz}.

In conclusion, we described the instability which occurs in a two-mode
photon system in the regime where the internal coupling is stronger than the
photon frequency. The ground state of the system is the vacuum below the
critical point, while in the other region a finite population of photons
appears. In the presence of a classical electron current, a
``superfluid-like'' phase manifests itself and gives rise to some
interesting effects which could experimentally observed, such as the
generation of an amount of entanglement in a time-evolution of factorized
coherent states independently from the incoming radiation amplitude.
Even if a zero-dimensional system a superfluid phase cannot be reached, by keeping the SBF to a finite value, we predict a short-time superfluid-like behavior.
We thank S. Paganelli for continuous and very useful discussions. S. Solimeno, P. Mataloni, and F. Illuminati are also acknowledged.

\end{document}